\documentclass[conference, 10pt, letterpaper]{IEEEtran}

\usepackage[utf8]{inputenc}
\usepackage{amsmath}
\usepackage{amsfonts}

\usepackage{amsthm}

\usepackage{float}

\usepackage{graphicx}
\graphicspath{{Figures/}}
\usepackage{epstopdf}

\usepackage{epsfig}
\usepackage{url}
\usepackage{comment}
\usepackage{cite}
\usepackage{graphics}
\usepackage{soul} 
\usepackage{multirow}

\usepackage{todonotes} 

\usepackage{graphicx, array, blindtext} 

\usepackage{graphicx}

\usepackage{wrapfig}

\usepackage{algorithm}  
\usepackage{algorithmic}

\usepackage{listings}
\usepackage{psfrag,wrapfig}

\usepackage{enumerate}
\usepackage{paralist} 
\usepackage{subfiles} 

\usepackage{array}
\usepackage{ragged2e}
\newcolumntype{P}[1]{>{\RaggedRight\hspace{0pt}}p{#1}}

\usepackage{multirow}

\newcommand{\eg}{{\it e.g.,}\xspace}
\newcommand{\viz}{{\it viz.,}\xspace}

\newcommand{\ie}{{\it i.e.,}\xspace}
\newcommand{\etc}{{\it etc.}}
\newcommand{\ci}{{\it (i) }}
\newcommand{\cii}{{\it (ii) }}

\newcommand{\ca}{{\it (a) }}
\newcommand{\cb}{{\it (b) }}

\newcounter{myoptimizationproblemctr}
\newcommand{\coolnameplus}{HYDRA\xspace}

\newcommand{\naive}{SingleCore\xspace}

\IEEEoverridecommandlockouts
\begin{document}

\title{A Design-Space Exploration for Allocating Security Tasks in Multicore Real-Time Systems
	}

\author{\IEEEauthorblockN{Monowar Hasan\IEEEauthorrefmark{1}, Sibin Mohan\IEEEauthorrefmark{1}, Rodolfo Pellizzoni\IEEEauthorrefmark{2} and Rakesh B. Bobba\IEEEauthorrefmark{3}} \IEEEauthorblockA{\IEEEauthorrefmark{1}Dept. of Computer Science, University of Illinois at Urbana-Champaign, Urbana, IL, USA}
\IEEEauthorblockA{\IEEEauthorrefmark{2}Dept. of Electrical and Computer Engineering, University of Waterloo, Ontario, Canada}
\IEEEauthorblockA{\IEEEauthorrefmark{3}School of Electrical Engineering and Computer Science, Oregon State University, Corvallis, OR, USA}
Email: \{\IEEEauthorrefmark{1}mhasan11, \IEEEauthorrefmark{1}sibin\}@illinois.edu,
\IEEEauthorrefmark{2}rodolfo.pellizzoni@uwaterloo.ca,
\IEEEauthorrefmark{3}rakesh.bobba@oregonstate.edu
}

\maketitle

\thispagestyle{plain}
\pagestyle{plain}


\begin{abstract}


The increased
capabilities of modern real-time systems (RTS) expose them to various security threats. Recently, frameworks that integrate security tasks without perturbing the real-time tasks have been proposed, but they only target single core systems. However, modern RTS are migrating towards multicore platforms. This makes the problem of integrating security mechanisms more complex, as designers now have multiple choices for where to allocate the security tasks. In this paper we propose {\em \coolnameplus}, a design space exploration algorithm that finds an allocation of security tasks
for multicore RTS using the concept of \textit{opportunistic execution}. \coolnameplus allows security tasks to operate with existing real-time tasks \textit{without} perturbing system parameters or normal execution patterns, while still meeting the desired monitoring frequency for intrusion detection. Our evaluation uses a representative real-time control system (along with synthetic task sets for a broader exploration) to illustrate the efficacy of \coolnameplus.

\vspace*{1em}

\end{abstract}

\section{Introduction}
\label{sec::intro}

Real-time systems (RTS) rely on a variety of inputs for their correct operation and have to meet stringent \textit{safety} and \textit{timing} requirements. The drive towards remote monitoring and control, increased connectivity through unreliable media such as the Internet and the use of component-based subsystems from different vendors are exposing modern RTS a multitude of threats. A successful attack on systems with real-time properties can have disastrous effects -- from loss of human life to damage to the environment and/or hard to replace equipment. A number of high-profile attacks on real systems, (\eg denial-of-service (DoS) attacks from Internet-of-Things devices \cite{ddos_iot_camera}, Stuxnet \cite{stuxnet}, 
 BlackEnergy \cite{Ukraine16}, \etc) have shown that the threat is real. Hence it is essential to retrofit existing critical RTS with detection, survival and recovery mechanisms. 
 
As the use of multicore platforms in safety-critical real-time systems is increasingly becoming common, the focus of this work is on integrating or retrofitting security mechanisms into \textit{multicore RTS}. It is not straightforward to retrofit RTS with security 
mechanisms that were developed for more general purpose computing scenarios since, security mechanisms have to \ca \textit{co-exist} with the real-time tasks in the system and \cb operate \textit{without} impacting the \textit{timing} and \textit{safety} constraints of the control logic. 
Further, it {\em may not be feasible to adjust the parameters (such as run-times, period, and execution order, \etc) of real-time tasks to accommodate security tasks.} This creates an apparent tension between security and real-time requirements. Unlike single core systems, integrating security into multicore platforms is more challenging since designers have multiple choices across cores to retrofit security mechanisms. For instance, is it better to \textit{dedicate a core} to all the security tasks or is it better to \textit{spread them out} (in conjunction with the real-time tasks) and if so, to \textit{which} cores? 


Our goal is to improve the security posture of multicore RTS by integrating security mechanisms \textit{without} violating real-time constraints. Security mechanism could include protection, detection or response mechanisms, depending on the system requirements -- for instance, a sensor correlation task (to detect sensor manipulation) or an intrusion detection task. As an illustrative example, consider the open source intrusion detection mechanisms Tripwire \cite{tripwire} and Bro \cite{bro} that detect integrity violations in the host and at the network level, respectively\footnote{We use Tripwire and Bro as examples of security applications to be integrated into multicore RTS -- the ideas presented here apply more broadly.}. The default configurations of Tripwire and Bro contain several tasks (see Table~\ref{table:rtos}). 

\begin{table}[!htb]
\vspace{-\baselineskip}
\caption{Illustration of Security Tasks\textsuperscript{*}}
\label{table:rtos}
\centering
\begin{tabular}{P{3.38cm} P{4.50cm}}
\multicolumn{2}{p{0.95\columnwidth}}{\textsuperscript{*}\footnotesize{The corresponding application for each of the security tasks is specified in the parenthesis -- TR: Tripwire, BR: Bro.}} \\
\hline
Task & Function\\
\hline
\hline 
Check own binary of the security routine (TR) & Compare the hash value of the application binary (\eg $\mathtt{/usr/sbin/tripwire}$, $\mathtt{/usr/local/bro/bin}$, \etc) \\
Check executables (TR)  & Check hash of the file-system binary ($\mathtt{/bin}$, $\mathtt{/sbin}$)  \\
Check critical libraries (TR) & Check library hashes ($\mathtt{/lib}$)    \\
Check device and kernel (TR)  & Check hash of peripherals and kernel information in $\mathtt{/dev}$ and $\mathtt{/proc}$\\
Check config files (TR) & Check configuration hashes ($\mathtt{/etc}$) \\
Monitor network traffic (BR)  & Scan network interface (\eg $\mathtt{en0}$) \\ 
\hline 
\end{tabular}
\vspace{-0.5em}
\end{table}

We propose to incorporate security mechanisms into a multicore setup by implementing them as separate \textit{sporadic tasks}. The challenge then, is to determine the \textit{right periods} (\viz minimum inter-execution time) and \textit{core assignment} for the security tasks. It is not trivial to determine the execution frequency and core assignment of security tasks (\ie what security tasks will execute on \textit{which core} and with \textit{what frequency}). For instance, some critical security routines may be required to execute more frequently than others. However, if the frequency of execution is too high then it will use too much of the processor time and lower the system utilization for real-time tasks. Hence, the security mechanism itself might prove to be a hindrance to the system and reduce the overall functionality or worse, negatively impact the system safety. In contrast, if the period is too long, the security task may not always detect violations since attacks could be launched between two instances of the security task. 
Existing work that integrates security in RTS either focuses on single core systems \cite{xie2007improving,lin2009static,zhang2013design, jiang2013optimization, mhasan_rtss16,sibin_RT_security_journal} and/or require modification of system parameters \cite{xie2007improving,lin2009static,zhang2013design, jiang2013optimization,sibin_RT_security_journal,securecore} and thus are not applicable for systems where it is harder to change the real-time task parameters.

In this paper we introduce \coolnameplus\footnote{In Greek mythology Hydra is a serpent with multiple heads. We refer to our scheme as \coolnameplus since we are trying to maximize the potential across multiple `heads' (cores).}, a scheme for multicore RTS that finds a suitable assignment of security tasks in order to ensure that they can execute with a frequency close to what a designer expects. The main contributions of this work can be summarized as follows: 

\begin{itemize}
\item Integrating security mechanisms in a multicore setup where changing existing real-time task parameters is not an option. 

\item A mathematical model to jointly obtain the assignment of security tasks to respective cores with  execution frequency close to the desired values (Section \ref{subsec:opt_formulation}). 

\item An iterative scheme, \coolnameplus, that jointly finds the assignment and period of the security tasks (Section \ref{subsec:algo}).

\item Comparisons of \coolnameplus with \ci assigning all security tasks to a single dedicated core and \cii an `optimal' multicore allocation scheme (Section~\ref{sec:result}).

\end{itemize}

We evaluate \coolnameplus with synthetic workloads as well as a representative real-time control system and security applications (Section \ref{sec:result}).

\section{System and Security Model}

\subsection{Real-time Tasks}

Let us consider a multicore platform comprised of $M$ identical
cores denoted by the set $\mathcal{M}= \lbrace \pi_1, \pi_2, \cdots, \pi_M \rbrace$ where we schedule a set $\Gamma_R = \lbrace \tau_1, \tau_2, \cdots ,\tau_{N_R} \rbrace$ of $N_R$ independent sporadic real-time tasks.  Each real-time task $\tau_r \in \Gamma_R$ is characterized by the tuple $(C_r, T_r, D_r)$ where $C_r$ is the worst-case execution time (WCET), $T_r$ is the minimum separation (\eg period) between two successive invocations and $D_r$ is the relative deadline. In this work, we consider \textit{partitioned
fixed-priority preemptive scheduling} \cite{mutiprocessor_survey} since \ca it does not introduce task migration costs and \cb it is widely supported in many commercial and open-source real-time
OSs (\eg QNX, OKL4, real-time Linux, \etc). We assume that real-time task priorities are distinct and assigned according to rate monotonic (RM) \cite{Liu_n_Layland1973} order. We also assume that tasks have implicit deadline, \eg $D_r = T_r, \forall \tau_r \in \Gamma_R$.

We assume that real-time tasks are \textit{schedulable} 
and assigned to the cores using existing multicore task partitioning algorithm \cite{mutiprocessor_survey}. Since the taskset is schedulable, the following necessary condition will hold \cite{baruah2005partitioned}:
\begin{equation} \label{eq:rt-sched}
\sum_{\tau_r \in \Gamma_R} \mathsf{DBF}(\tau_r, t) \leq Mt, \quad \forall t>0
\end{equation}
where the \textit{demand bound function} $\mathsf{DBF}(\tau_r, t)$ computes the cumulative maximum execution requirements of the real-time task $\tau_r$ and it is defined as follows: 
$\mathsf{DBF}(\tau_r, t) = \max \left( 0, \left( \left \lfloor \frac{t- D_r}{T_r} \right\rfloor + 1 \right) C_r \right).$

\subsection{Threat Model}

In this work we consider a generic threat model where a malicious adversary can use various techniques to attack
the RTS. For example, the adversary might intercept the information over the communication channel,
forge messages or
prevent normal requests from being processed. The adversary can also attack services
within the OS, say, could compromise the file system resulting in corrupted
information or could delay the delivery of control
commands that may cause some tasks to miss deadline. Other than trying to aggressively crash the system, the adversary may utilize side-channels to monitor the system behavior and infer certain system information (\eg user tasks, thermal profiles, cache information, \etc) that eventually leads to the attacker actively taking control of the system.  Our focus is on threats that can be dealt with by integrating additional security tasks. The addition of such tasks may necessitate changing the schedule of real-time tasks as was the case in earlier work \cite{sibin_RT_security_journal,lin2009static,xie2007improving, zhang2013design, jiang2013optimization}. In this work we focus on situations where added security tasks are not allowed to impact the schedule of existing real-time tasks as is often the case when integrating security into existing multicore systems.

\subsection{Security Tasks}

Since our goal is to \textit{ensure security without any modification of real-time task parameters}, we propose to integrate security tasks as independent sporadic tasks. Let us consider additional $N_S$ security tasks denoted by the set $\Gamma_S = \lbrace \tau_1, \tau_2, \cdots ,\tau_{N_S} \rbrace$. We follow the sporadic security task model \cite{mhasan_rtss16} and characterize each security task $\tau_s$ by the tuple $(C_s, T_s^{des}, T_s^{max})$ where $C_s$ is the WCET, $T_s^{des}$ is
the best  period (minimum inter-arrival time) between successive releases (\ie
$F_s^{des} = \frac{1}{T_s^{des}}$
is the desired frequency for $\tau_s$ effective security monitoring and/or intrusion detection) and $T_i^{max}$ is the maximum period beyond which security monitoring will not be effective. We assume that periods for security tasks are assigned based on the desired monitoring frequency. Hence $pri(\tau_{s_1}) > pri(\tau_{s_2})$ if $T_{s_1}^{max} < T_{s_2}^{max}$ where $pri(\tau_{i})$ denotes the priority of $\tau_i$. Security tasks also have implicit deadlines (\eg they are required to complete execution before its period).

One fundamental problem while integrating security mechanisms is to determine \textit{which security tasks will be assigned to which core and executed when}.
Although security tasks can execute in any of the $M$ available cores and any period $T_s^{des} \leq T_s \leq T_s^{max}$ is acceptable, the actual task-to-core assignment and the periods of the security tasks are not known apriori. The goal of \coolnameplus therefore is to jointly find the core-to-task assignment and suitable periods for security tasks.

\section{Assignment of Security Tasks with Period Adaptation}


One way to integrate security mechanisms into existing systems without perturbing real-time task behavior is to execute security tasks with the \textit{lowest priority} as compared to the real-time tasks \cite{mhasan_rtss16}. Thus security tasks will execute \textit{opportunistically} in the slack time (\eg when other real-time tasks are not running). As mentioned earlier, actual periods
of the security tasks are not known and we need to \textit{adapt} the
periods within acceptable ranges to optimize the trade-offs
between schedulability and defense against intrusions. We measure the security of the system by means of the
\textit{achievable periodic monitoring} and
our goal is to minimize the perturbation between the achievable (unknown)
period $T_s$ and the given desired period $T_s^{des}$ for all security tasks $\tau_s \in \Gamma_S$. Therefore we consider the following metric \cite{mhasan_rtss16}:
\begin{equation} \label{eq:eta}
\eta_s = \frac{T_s^{des}}{T_s}
\end{equation}
that denotes the \textit{tightness} of periodic monitoring (\eg how close the period of the security task is to the desired period) and bounded by $\frac{T_s^{des}}{T_s^{max}} \leq \eta_s \leq 1$. As mentioned
earlier, if the interval between consecutive monitoring events
is too large, the adversary may remain undetected and harm
the system between two invocations of the security task. On the other hand, very frequent execution of security tasks
may impact the schedulability of the system (due to higher utilization). The
metric in Eq. (\ref{eq:eta}) allows us to measure how close the security tasks are able to get to their desired monitoring frequencies.


Note that arbitrarily setting $T_s = T_s^{des}$ for all (or some) security tasks $\tau_s \in \Gamma_S$ may lead to the system becoming \textit{unschedulable} since low-priority security tasks may miss deadlines due to interference from higher priority tasks. Also exhaustively finding all possible acceptable periods for the security tasks for all available cores is not feasible. It will cause an exponential blow-up as numbers of tasks and cores increase. For instance  for a given taskset $\Gamma_S$, there is a total of $|\mathcal{M} \times \Gamma_s|$ assignments possible  (where $A \times B = \lbrace (a,b) ~|~ a\in A \wedge b\in B \rbrace$ and $|\cdot|$ denotes set cardinality) and for each combination the period for each security task $\tau_s \in \Gamma_S$ can be any value within the range $[T_s^{des}, T_s^{max}]$. In order to address this combinatorial problem we obtain the periods of the security tasks by framing it as an optimization problem.

\subsection{Formulation as an Optimization Problem} \label{subsec:opt_formulation}

\subsubsection{Objective Function and Bounds on Period}

Let us consider the vector $\mathbf{X} = [x_s^m]_{\forall \tau_s \in \Gamma_S, \forall \pi_m \in \mathcal{M}}^\mathrm{T}$ where 
$x_s^m = 1$ if $\tau_s$ is assigned to $\pi_m$ and $0$ otherwise.
Recall that our goal is to find a task assignment
that minimizes the difference between achievable and desired periods (\eg maximize the tightness) for all the security tasks. Hence we define the following objective function:
\begin{equation} \label{eq:objective_function}
\underset{\mathbf{X}, \mathbf{T}}{\operatorname{max}} ~~ \sum_{\pi_m \in \mathcal{M}} \sum_{\tau_s \in \Gamma_S} x_s^m \omega_s \eta_s = \sum_{\pi_m \in \mathcal{M}} \sum_{\tau_s \in \Gamma_S} x_s^m \omega_s  \frac{T_s^{des}}{T_s}
\end{equation}
where $\mathbf{T} = [T_s]_{\forall \tau_s \in \Gamma_s}^{\mathrm{T}}$ is the (unknown) period vector that needs to be determined and $\omega_s$ reflects the priority (higher priority tasks would have large $\omega_s$). Besides, in order to satisfy the frequency of periodic monitoring, the security task needs to satisfy the following constraint:
\begin{equation} \label{eq:period_bound}
T_s^{des} \leq T_s \leq T_s^{max}, ~\forall \tau_s \in \Gamma_s.
\end{equation}
Finally, each security task must be assigned to exactly one core:
$\sum\limits_{\pi_m \in \mathcal{M}} x_s^m = 1,~~ \forall \tau_s \in \Gamma_s.$

\subsubsection{Schedulability Constraint}

Since the security tasks are executed with a priority lower that all real-time tasks, they will suffer interference from all real-time and high priority security tasks executing in the same core. Let $hp_S(\tau_s) \subset \Gamma_S$ denote the set of security tasks with a higher priority than $\tau_s$. The worst-case release pattern of $\tau_s$ occurs when
$\tau_s$ and all high-priority tasks are released simultaneously \cite{res_time_rts}. Using response time analysis \cite{mn_gp} we can determine an upper bound to the interference experienced by $\tau_s$ for a given core $\pi_m$ as follows:
\begin{equation} \label{eq:intf}
I_s^m = \!\! \sum_{\tau_r \in \Gamma_R } \mathbb{I}_r^m \left( 1 + \frac{T_s}{T_r} \right) C_r + \!\!\!\!\! \sum_{\tau_h \in hp_S(\tau_s)} \!\!\!\! x_h^m \left( 1 + \frac{T_s}{T_h} \right) C_h
\end{equation}
where $\mathbb{I}_r^m = 1$ if the real-time task $\tau_r$ is partitioned to core $\pi_m$ and $0$ otherwise. 

The first and second term in Eq. (\ref{eq:intf}) represent the amount of interference from real-time and high-priority security tasks, respectively. Note that the assignment of real-time tasks to cores is known by assumption. In order to ensure that each security task $\tau_s$ will complete its execution before its deadline on its assigned core, the following constraint needs to be satisfied:
\begin{equation} \label{eq:sched}
C_s + I_s^m \leq T_s, ~~\forall \tau_s \in \Gamma_s, \forall \pi_m \in \mathcal{M}:  x_s^m = 1.
\end{equation}

The variables $\mathbf{X}$ and $\mathbf{T}$ in the above formulation turn the problem into a \textit{non-linear combinatorial optimization problem} that is NP-hard. We therefore propose an iterative algorithm \coolnameplus that jointly finds the security tasks' period and core assignment.




\subsection{Algorithm Development}
\label{subsec:algo}

As mentioned earlier,  jointly finding the security task assignment and periods is an NP-hard problem. Even for fixed periods, finding  the assignment for security tasks turns the problem to a bin-packing problem that is known to be NP-hard \cite{baruah2005partitioned}. 
Existing partitioning heuristics (\eg first-fit, best-fit, \etc) \cite{mutiprocessor_survey} are \textit{not} directly applicable in our context since the real-time requirements  (\eg  minimize the number of cores so that all real-time tasks can meet deadlines) are often different from the security requirements (\eg execute security tasks more often to improve intrusion detection rate without violating real-time constraints). 

For a given task $\tau_s$ and allocation vector $\mathbf{X}$, let us rewrite the optimization problem as follows: 
\begin{equation}\label{eq:opt_period}
\underset{T_s}{\operatorname{max}}~  \eta_s, ~\text{Subject to:}~ T_s^{des} \leq T_s \leq T_s^{max},~ C_s + I_s^m \leq T_s.
\end{equation}
Notice that for a given assignment $\mathbf{X}$ (see Algorithm \ref{alg:multicore_sec}), the period $T_s$ is the only variable (when the $T_h, \forall \tau_h \in hp_S(\tau_s)$ is known) in $I_s^m$ (see Eq. (\ref{eq:intf})). 
Although the period adaptation problem in Eq. (\ref{eq:opt_period}) is a  constraint optimization problem it can be transformed into a convex optimization problem (that is solvable in polynomial time). 
For details of this reformulation we refer the readers to the Appendix.




\renewcommand{\algorithmicforall}{\textbf{for each}}
\renewcommand\algorithmiccomment[1]{%
 {\it /* {#1} */} %
}
\renewcommand{\algorithmicrequire}{\textbf{Input:}}
    \renewcommand{\algorithmicensure}{\textbf{Output:}}
    
		\begin{algorithm}
			\begin{algorithmic}[1]
				\begin{footnotesize}
				\REQUIRE Input taskset $\Gamma =\lbrace  \Gamma_R \cup \Gamma_S \rbrace$ and the partition of real-time tasks $\mathbb{I} = [\mathbb{I}_r^m]_{\forall \tau_r \in \Gamma_R, \forall \pi_m \in \mathcal{M}}^{\mathrm{T}}$
   
   \ENSURE The security task allocation $\mathbf{X}=[x_s^m]_{\forall \tau_s \in \Gamma_S, \forall \pi_m \in \mathcal{M}}^\mathrm{T}$ and periods $\mathbf{T} = [T_s]_{\forall \tau_s \in \Gamma_S}^\mathrm{T}$, if the taskset is schedulable, $\mathsf{Unschedulable}$ otherwise.
					\vspace{0.4em}
            	     \STATE Initialize $x_s^m :=0, ~ \forall \tau_s \in \Gamma_S, ~\forall \pi_m \in \mathcal{M} $   
                    \FORALL{security task $\tau_s \in \Gamma_S$ (from higher to lower priority)}
                      \FORALL{core $\pi_m \in \mathcal{M}$}
                      \STATE Solve the  optimization problem in Eq. (\ref{eq:opt_period})
                      \ENDFOR
                      \STATE Let $\mathcal{M}_s^\prime \subseteq \mathcal{M}$ is the set of core(s) for which the optimization problem is feasible
                      \IF{$\mathcal{M}_s^\prime = \emptyset$}
                      \STATE \COMMENT{Unable to find suitable period for $\tau_s$}
                      \RETURN \! $\mathsf{Unschedulable}$
                      \ENDIF
                      
                      \STATE Find the core $\pi_{m^*} = \underset{\pi_m \in \mathcal{M}_s^\prime}{\operatorname{argmax}} ~\eta_s^m$ where $\eta_s^m$ is the tightness of $\tau_s$ obtained for $\pi_m$
                      
                      \STATE Set $x_s^{m^*} := 1$ ~~\COMMENT{Assign $\tau_s$ to $\pi_{m^*}$}
                      \STATE Update period $T_s := T_s^{m^*}$ where $T_s^{m^*}$ is the period obtained by solving optimization for $\pi_{m^*}$
                    \ENDFOR
                    
                   	\RETURN $(\mathbf{X},  \mathbf{T})$ ~~\COMMENT{Return the allocation vector and periods}
                    \end{footnotesize}
				\end{algorithmic}
                
                \caption{\coolnameplus: Task Allocation and Period Adaptation}
			\label{alg:multicore_sec}
                \end{algorithm}

The proposed \coolnameplus algorithm (summarized in Algorithm \ref{alg:multicore_sec}) works as follows. We start with the highest priority security task $\tau_s$ and try to obtain the best period for all available core $\pi_m \in \mathcal{M}$ by solving the period adaptation problem introduced in Eq. (\ref{eq:opt_period}) (Line 4). If there exists a set of cores $\mathcal{M}_s^\prime \subseteq \mathcal{M}$ for which the optimization problem is feasible (\eg an optimal period is obtained satisfying the real-time constraint) we pick the core $\pi_{m^*} \in \mathcal{M}_s^\prime$ that gives the maximum tightness (Line 11) and allocate the security task to core $\pi_{m^*}$ (Line 12). This will ensure that the more critical security tasks will get a period close to the desired one. We repeat this process for all security tasks to jointly obtain the assignment and periods. If for any security task $\tau_j$ the set of available cores $\mathcal{M}_j^\prime$ is empty (\eg the optimization problem is infeasible) we return the taskset as \textit{unschedulable} (Line 9) since it is not possible to find any suitable core with given taskset parameters. This unschedulability result will provide hints to the designers to update the parameters of security tasks (and/or the real-time tasks, if possible) in order to integrate security for the target system.


\section{Evaluation} 
\label{sec:result}

We evaluated HYDRA along two fronts: 
\ci on parameters derived from a real UAV control system (Section \ref{sec:exp_case_study}) and
\cii synthetically generated tasksets to explore the design space (Section \ref{sec:exp_synthetic}). Recall from Section \ref{sec::intro} that our goal is to explore the possible
ways in which security could be integrated in multicore-based real-time platforms. The HYDRA mechanism presented in this paper assumes that the real
time tasks are distributed across {\em all} available cores. Another design
choice available is to allocate a dedicated core for security while the 
real-time tasks are assigned to the remaining cores.
In this Section, we compare  \coolnameplus to this alternate mechanism for security task 
allocation -- that we refer to henceforth as the ``\naive'' 
allocation mechanism.
Given the taskset is schedulable, one of the benefits of the \naive scheme is that there is no requirement for assigning security tasks. While evaluating \naive, all the real-time tasks are partitioned into $M-1$ cores leaving the other core for security tasks.
Notice that in the \naive scheme security tasks do not suffer any interference from real-time tasks (\eg the first term in Eq. (\ref{eq:intf}) is zero). For a given assigned core $\pi_m$, the decision variable $x_s^m$ is known for all $\tau_s$ and the optimization problem can solved using an approach similar to the one described in the Appendix. 

\subsection{Case-study with a UAV Control System and Security Applications} \label{sec:exp_case_study}

The goal of this experiment was to observe the runtime behavior of \coolnameplus. For a real-time application, we considered a UAV control system
\cite{uav_flight_control}. It includes following real-time tasks (refer to earlier work \cite[Tab. 1]{uav_flight_control} for the task parameters):
\textit{Guidance} (selects the reference trajectory), \textit{Slow  and Fast navigation} (read sensor values according to the required update frequency), \textit{Controller} (executes the closed-loop control functions), \textit{Missile control} (fires missile) and 
\textit{Reconnaissance} (collects sensitive information and send data to the control center). For the security application, we considered Tripwire \cite{tripwire}
 and Bro \cite{bro}
that detects integrity violations in the system both at host and network level, respectively (refer to Table \ref{table:rtos}). We executed the security tasks on an 1 GHz ARM Cortex-A8 processor with Xenomai 2.6 \cite{xenomai} patched real-time Linux kernel (version 3.8.13-r72) and used ARM cycle counter registers (\eg CCNT) to obtain the timing parameters (\eg WCET). We used $\mathtt{GPkit}$ \cite{gpkit} library and $\mathtt{CVXOPT}$ \cite{cvxopt} solver to obtain the periods.

\begin{figure}[!t]
\vspace*{-\baselineskip}
\centering
\advance\leftskip-0.8em
\includegraphics[scale=0.40]{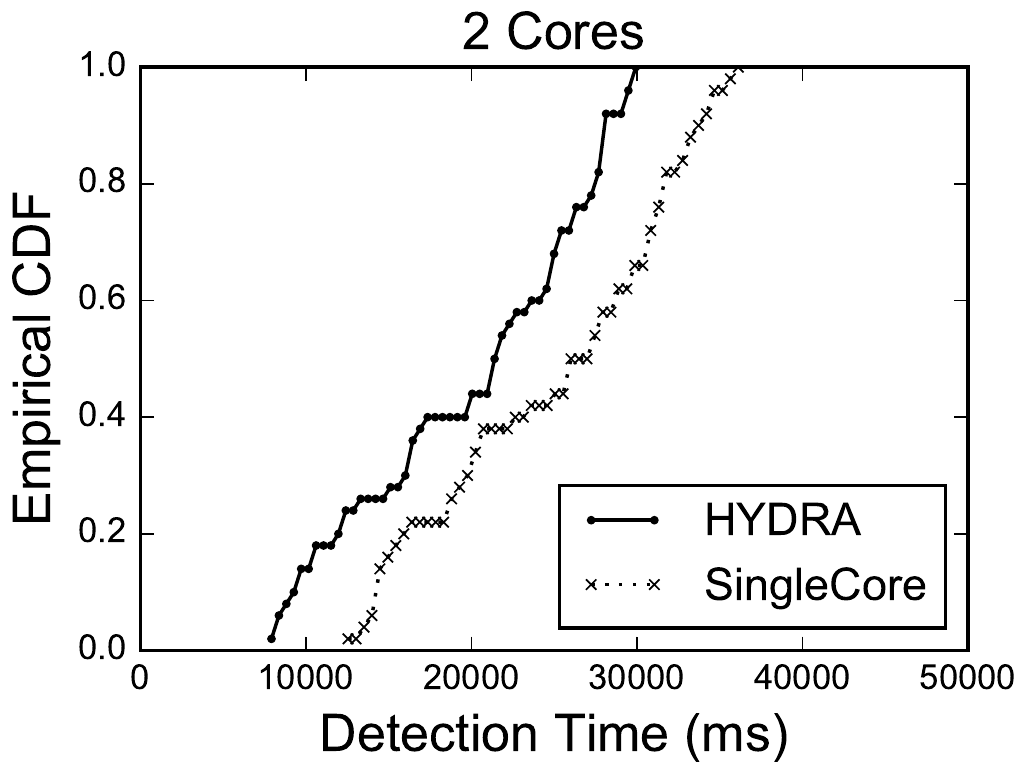}
\includegraphics[scale=0.40]{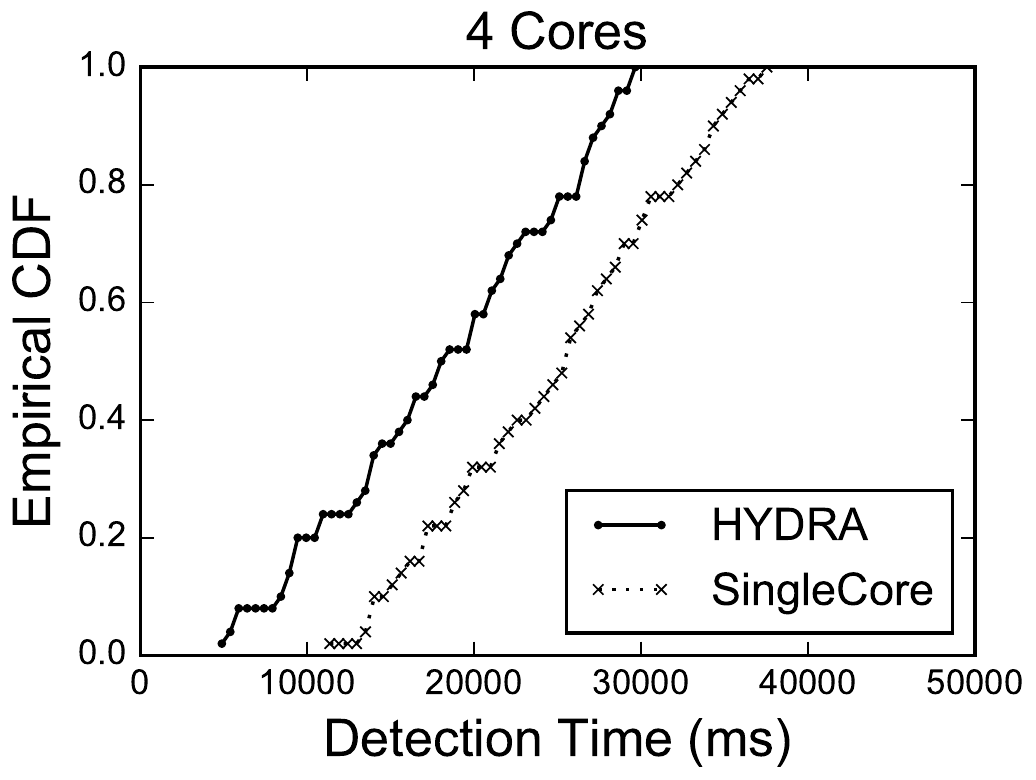}
\includegraphics[scale=0.40]{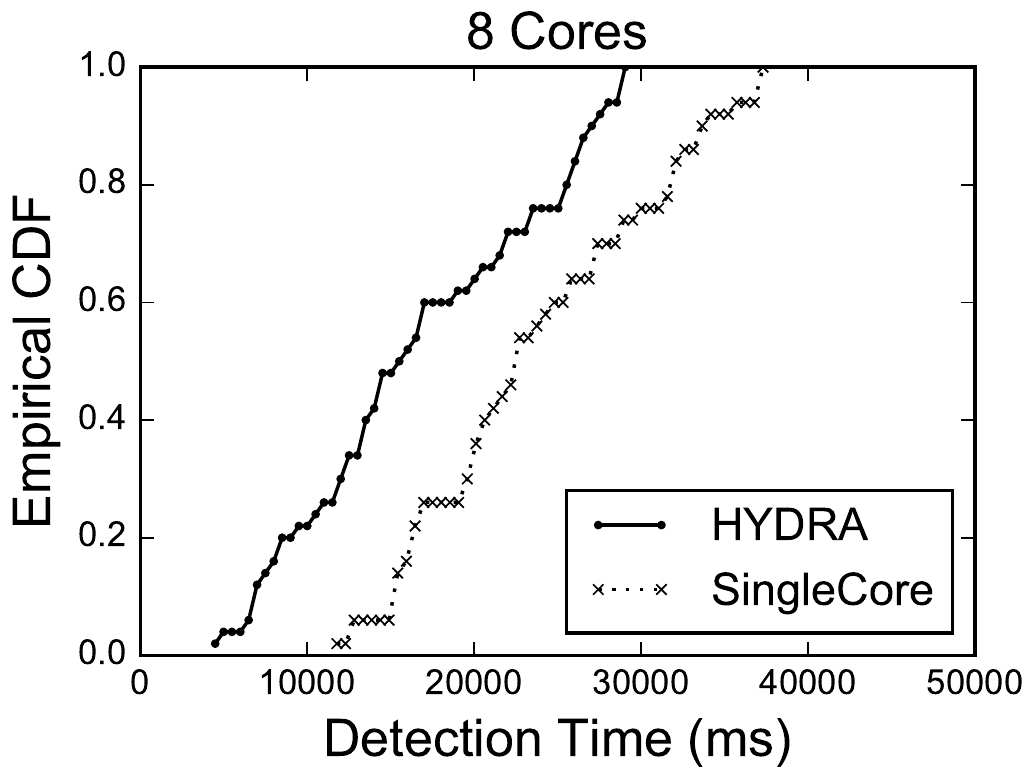}
\caption{\coolnameplus vs. \naive: empirical CDF of intrusion detection time. The empirical CDF is defined as
$
\widehat{F}_\alpha(\jmath) = \frac{1}{\alpha} \displaystyle \sum\limits_{i=1}^\alpha \mathbb{I}_{[\zeta_i \leq \jmath]},
$
where $\alpha$ is the total number of experimental observations, ${\zeta}_i$ is time to detect the attack in at the $i$-th experimental observation and $\jmath$  represents the $x$-axis values (\eg detection time).
The indicator function $\mathbb{I}_{[\cdot]}$ outputs $1$ if the condition $[\cdot]$ is satisfied and $0$ otherwise.} 
\label{fig:cs}
\vspace{-0.25in}
\end{figure}

The work-flow of our experiment was as follows. For each of the trials, we observed the schedule for $500 ~ \!\rm{s}$ and during any random time of execution we triggered synthetic attacks\footnote{Our goal here is to analyze the security from the scheduling perspective. Thus instead assuming any specific intrusion (or detection capabilities of security tasks), \coolnameplus allows designer to integrate any security mechanism required to defend targeted attack surfaces.} (\eg that corrupts the file system and network packets). We assume that the intrusions are correctly detected by the security tasks (\eg there is no false positive/negative errors) and measured the empirical CDF of worst-case detection time. From Fig. \ref{fig:cs} we can observe that paralleling security tasks across cores leads to faster intrusion detection time for \coolnameplus (\eg higher empirical CDF). From our experiment we found that \textit{on average} \coolnameplus can provide $19.81\%$, 
$27.23\%$ and
$29.75\%$ faster detection rate for $2$, $4$ and $8$ core system, respectively. While \naive scheme does not experience any interference from real-time tasks, however, low priority security tasks can still suffer inference from high priority security tasks. Therefore running security tasks in a single core leads to higher periods and consequently poorer detection time.


\subsection{Experiment with Synthetic Tasksets} 
\label{sec:exp_synthetic}



We used parameters similar to those in related work
\cite{davis2015global, mhasan_rtss16}. We performed experiments for $M=\lbrace 2, 4, 8 \rbrace$ cores. Each taskset instance contained $[3M, 10M]$
real-time and $[2M, 5M]$ security tasks. Each real-time task had periods between $[10~\mathrm{ms}, 1000~\mathrm{ms}]$. The desired periods for the security tasks were selected from $[1000~\mathrm{ms}, 3000~\mathrm{ms}]$ and the maximum allowable period
is assumed to be $T_s^{max} = 10T_s^{des}, ~\forall \tau_s$. The real-time tasks are partitioned across multiple cores using a best-fit \cite{mutiprocessor_survey} strategy. 

In each experiment, the total
taskset utilization was varied from $0.025M$ to $0.975M$ with step size $0.025M$. For a given number of tasks and total system utilization, the utilization of individual tasks were generated from an unbiased set of utilization values using the Randfixedsum algorithm \cite{randfixedsum}. Total utilization of the security
tasks were set to be no more than 30\% of the real-time
tasks.  For each
utilization value, we randomly generated $250$ tasksets. In other words, for each core configuration a total of $39 \times 250 =9750 $ tasksets were tested. We only considered tasksets that satisfied the necessary condition in Eq. (\ref{eq:rt-sched}), as any taskset that fails the condition is trivially unschedulable.
\vspace{\baselineskip}





\begin{figure}[!t]
\vspace*{-\baselineskip}
\centering
\advance\leftskip-0.8em
\includegraphics[scale=0.40]{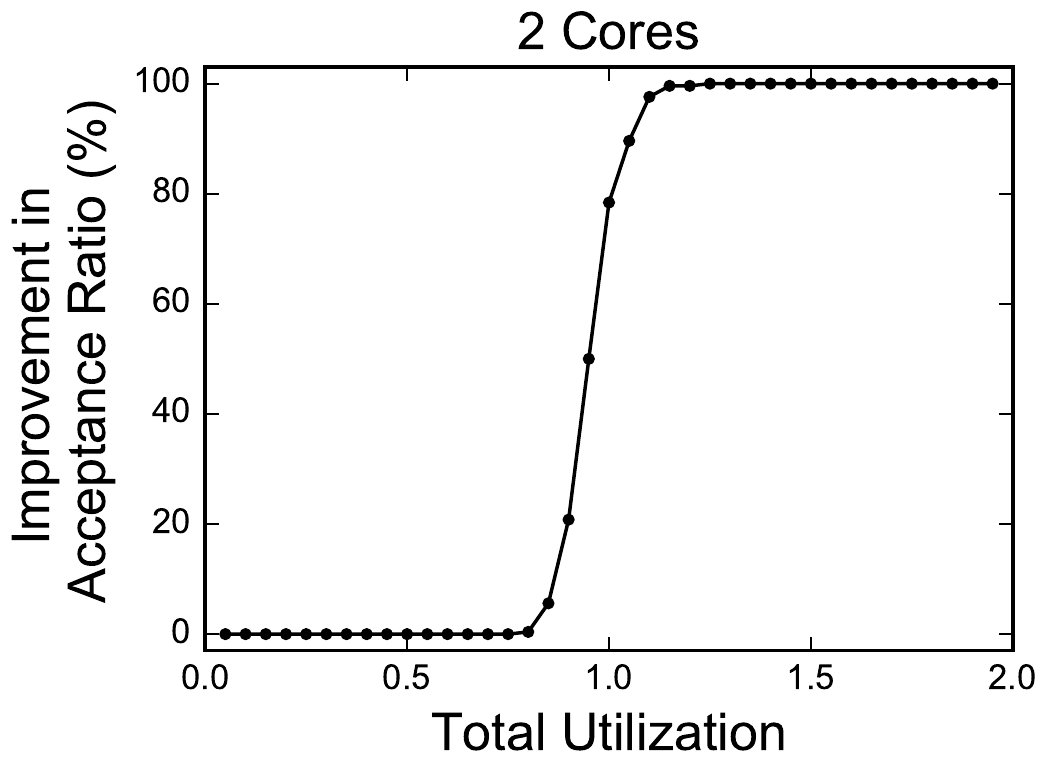}
\includegraphics[scale=0.40]{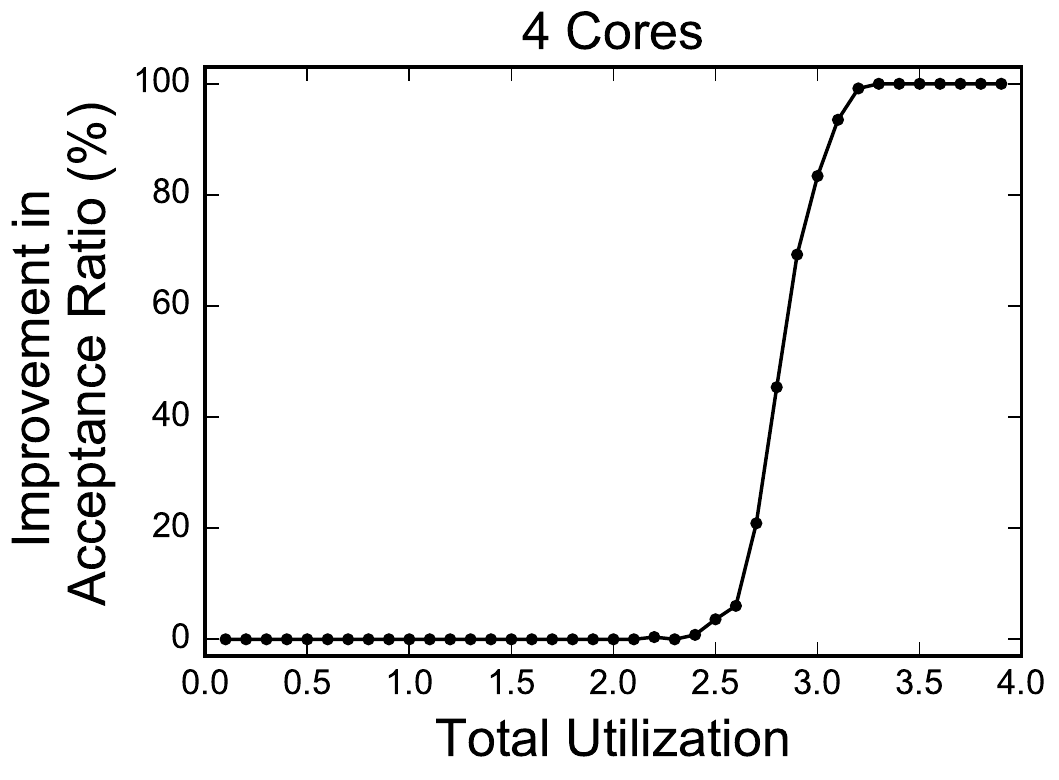}
\includegraphics[scale=0.40]{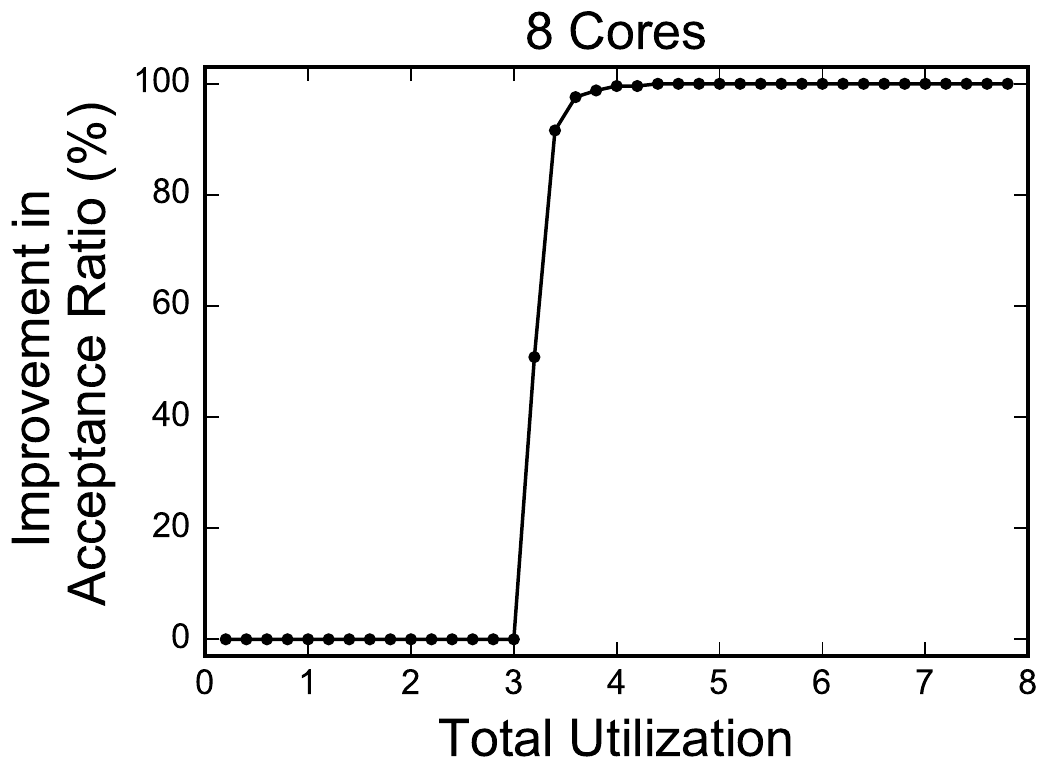}
\caption{The improvement in acceptance ratio for $2, 4,$ and $8$ core system. The improvement is given by $\frac{\delta_{\rm SingleCore}-\delta_{\rm HYDRA}}{\delta_{\rm SingleCore}} \times  100 \%$ where $\delta_{(\cdot)}$ is the acceptance ratio of scheme $(\cdot)$.} 
\label{fig:sched}
\vspace{-0.25in}
\end{figure}

\noindent {\em 1. Experiment with Core Assignment Schemes:} 
We compared \coolnameplus with \naive in terms of \textit{acceptance ratio}. The acceptance ratio is given by the number of \textit{schedulable} tasksets (\eg that satisfy all real-time constraints) over the generated ones. The x-axis in Fig. \ref{fig:sched} represents the total system utilization (\eg utilization of both real-time and security tasks). The y-axis represents improvement in acceptance ratio comparing \coolnameplus with \naive for different values of $M$. For lower utilization values both schemes have similar performance (\eg improvement is zero) since the system has enough slack to execute security tasks. However as we see from the figure, for higher utilization values \coolnameplus outperforms \naive~-- when all security tasks share a core, it causes more interference and reduces the overall schedulability (\eg unable to find any solution that respects all the real-time constraints\footnote{Note that security tasks also have real-time constraints.}).

\begin{figure}[t]
\vspace*{-\baselineskip}
\centering
\includegraphics[scale=0.45]{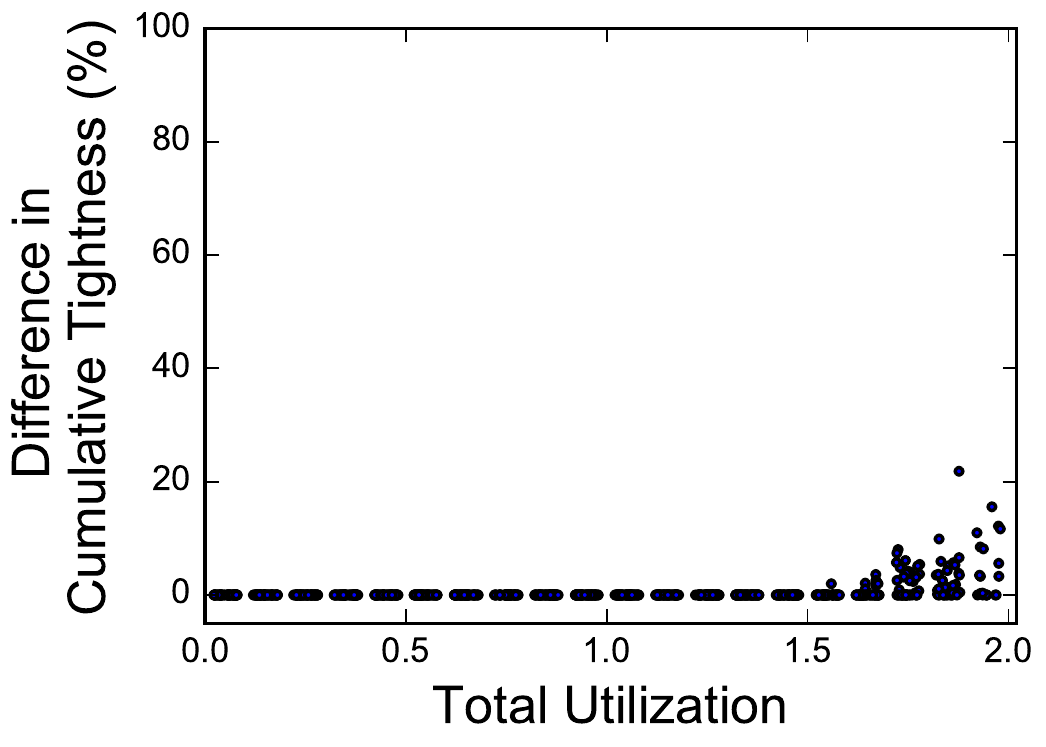}
\caption{Comparing \coolnameplus with optimal solution: we consider $M=2$ and $N_S \in [2, 6]$ with other parameters similar to that mentioned in Section \ref{sec:exp_synthetic}.} 
\label{fig:esearch}
\vspace{-0.25in}
\end{figure}

\vspace{\baselineskip}
\noindent {\em 2. Comparing with Optimal Multicore Assignment:}
The result of an empirical comparison of \coolnameplus with an optimal (exhaustive) solution is presented in Fig. \ref{fig:esearch} where we searched for all possible combinations for a small setup with $M = 2$ cores and up to $N_S = 6$ security tasks. To find the optimal solution, we test each of the $M^{N_S}$ possible assignments of security tasks to cores. For each assignment, we then determine the value of the period vector $\mathbf{T}$ that maximizes the cumulative tightness by solving a convex optimization problem in polynomial time (see Appendix). 

The x-axis of Fig. \ref{fig:esearch} represents total system utilization and y-axis is the difference in cumulative tightness (\eg $\Delta_{\eta} = \frac{\eta_{\rm OPT} - \eta_{\rm HYDRA}}{\eta_{\rm OPT}}\times 100 \%$) 
for \coolnameplus and the optimal solutions. As shown in the figure, for low to medium utilization cases, \coolnameplus's performance is similar to the optimal solution (\eg the difference is zero). However for higher utilizations performance degrades. This is because \coolnameplus follows an iterative best-fit strategy to find the periods (and assignment). Hence for higher utilization values the lower priority tasks may not get periods close to the desired values (and the cumulative tightness degrades). As we see from the figure, the degradation (in cumulative tightness) is no more than $22\%$ and that may be acceptable given the exponential computational complexity of finding an optimal solution.

\section{Discussion}

While we take a step towards developing a model for integrating security mechanisms into multicore RTS, our initial attempt can be extended into several directions. \coolnameplus statically partition the tasks to the cores. However in practice security tasks can also move across multiple cores if there is available slack at runtime (for faster detection and better schedulability). While there exists methods for global scheduling policy \cite{mutiprocessor_survey} where tasks can migrate across cores, casting real-time scheduling problems into RTS security domain requires further research. 

In this work we consider security tasks are independent and preemptive. However some critical security task may require non-preemptive execution to perform desired checking. In addition, depending on the actual implementations of the security routines, the scheduling framework may need to follow certain \textit{precedence constraints}. For example, in order to ensure that the security application itself has not been compromised, the security application's own binary may need to be examined first before checking the system binary files. These aspects will be explored in our future work.

\section{Related Work}

There has been some work \cite{xie2007improving, lin2009static, jiang2013optimization, zhang2013design} on reconciling the addition of security mechanisms into RTS that considered periodic task scheduling where each task requires a security service whose overhead varies according to the quantifiable level of the service. The issues regarding information leakage through storage channels also addressed in prior research \cite{sibin_RT_security_journal}.  All of the aforementioned work, however, only consider single core system and require modification of system parameters. A similar line of work \cite{mhasan_rtss16} to our exists where authors used the concept of hierarchical scheduling proposed to execute the security mechanisms with a lower priority than the real-time tasks for a single core system. Unlike prior work we focus on integrating security in multicore domain.


Although not in the context of security in RTS, there exists other work \cite{delay_period, davare2007period_can} in which
the authors statically assign the  periods for control tasks. While this previous work focused on single core systems and optimizing period of \textit{all} the tasks, our goal is to ensure security without violating timing constraints of the real-time tasks in a multicore setup. 

In contrast to proposed scheduler-level solution, recent work \cite{slack_cornell, securecore,mhasan_resecure16} on hardware/software architectural frameworks aim to protect multicore RTS against security vulnerabilities. Compared to our scheme that works for any $m$-core system, these preceding frameworks mainly focus on dual core architecture and require architectural modifications that may not be suitable for existing RTS. 

\section{Conclusion}
\label{sec::concl}

This paper presents an evaluation of a good heuristic mechanism (\coolnameplus) for assigning security
tasks into a multicore RTS. Engineers can now evaluate the design choices of such assignments
to improve the overall security (and hence, safety) of systems with real-time requirements. Since we provide comparisons of our
solution with two extremes -- an `optimal' assignment strategy and isolating all security tasks to a single core
-- we are able to provide valuable hints to designers on how to build security into such systems.


\bibliographystyle{IEEEtran}

\bibliography{references_short}



\appendix[Solution to the Period Adaptation Problem]


The period adaptation problem given in Section \ref{subsec:opt_formulation} is a constrained optimization problem and 
not very straightforward to solve. Therefore we reformulate the optimization problems as a geometric program (GP) \cite{GP_tutorial}. A nonlinear optimization problem can be solved by GP if the problem is formulated as follows \cite{GP_tutorial}:
\begin{align*}
\underset{\mathbf{Y}}{\operatorname{min}}~  f_0(\mathbf{y}),  
\text{~Subject to:~} 
 f_i(\mathbf{y}) &\leq  1, \quad i = 1, \cdots, z_p,  \text{~~and~~} \\
 g_i(\mathbf{y}) &=  1, \quad i = 1, \cdots, z_m
\end{align*}
where $\mathbf{y} = [y_1, y_2, \cdots, y_z]^{\mathrm{T}}$ denotes the vector of $z$ optimization variables. The functions $f_0(\mathbf{x}), f_1(\mathbf{y}), \cdots, f_{z_p}(\mathbf{y})$ are \textit{posynomial} and $g_1(\mathbf{y}), \cdots, g_{z_m}(\mathbf{y})$ are \textit{monomial} functions, respectively. A monomial function is  expressed as
$
g_i(\mathbf{y}) = c_i \prod\limits_{l = 1}^{L_i} y_l^{a_l},
$
where $c_i \in \mathbb{R}^+$ and $a_l \in \mathbb{R}$. 
A posynomial function (\ie the sum of the monomials) can be represented as
$
f_i(\mathbf{y}) = \sum\limits_{l=1}^{L_i} c_l y_1^{a_{1l}} y_2^{a_{2l}} \cdots y_z^{a_{1l}},
$
where $c_l \in \mathbb{R}^+$ and $a_{jl} \in \mathbb{R}$. 
We can maximize a non-zero posynomial objective function by minimizing its inverse. In addition, we can express the constraint $f(\cdot) < g(\cdot)$ as $\frac{f(\cdot)}{g(\cdot)} \leq 1$.

Based on above discussion we can rearrange the objective function as $\underset{T_s}{\operatorname{min}}~  {(T_s^{des})}^{-1}$. Likewise period bound constraint in Eq. (\ref{eq:period_bound}) can be represented as $T_s^{des} {T_s}^{-1} \leq 1$ and
 ${(T_s^{max})}^{-1} T_s \leq 1$, respectively. In addition, the schedulability constraint in Eq. (\ref{eq:sched}) can be rewritten as: 
$(C_s + I_s^m) T_s^{-1} \leq 1$ where 
$I_s^m = \!\!\! \sum\limits_{\tau_r \in \Gamma_R} \mathbb{I}_r^m (T_r + T_s) T_r^{-1}C_r + \!\!\!\!\!\!\! \sum\limits_{\tau_h \in hp_S(\tau_s)} \!\!\!\!\! x_h^m (T_h + T_s)T_h^{-1} C_h$.
                  
The above GP reformulation is not a convex optimization problem since the posynomials are not convex functions \cite{GP_tutorial}. However, by using logarithmic transformations (\eg representing $\tilde{T}_s = \log T_s$ and hence $T_s = e^{\tilde{T}_s}$, and replacing inequality constraints of the form $f_i(\cdot) \leq 1$ with $\log f_i(\cdot) \leq 0$), we can convert the above formulation into a convex optimization problem that can be solved using standard algorithms, such as \textit{interior-point} method in polynomial time \cite[Ch. 11]{boyd_book}.

\end{document}